\documentclass{amsart}
\def \iM {\mathcal M}
\def \iS {\mathcal S}

\def \iP {\mathcal P}
\def \vfi {\varphi}
\def \supp {{\rm supp}\,}
\newtheorem{lemma}{Lemma}[section]
\newtheorem{prop}{Proposition}[section]
\newtheorem{coro}{Corollary}[section]
\newtheorem{thm}{Theorem}[section]
\usepackage{amssymb}

\begin{document}

\title{ A construction of a nonparametric quantum information manifold.}
\author{Anna Jen\v cov\'a}
\maketitle
\centerline{\small Mathematical Institute, Slovak Academy of Sciences,}
\centerline{\small  \v Stef\'anikova
49, 814 73 Bratislava, Slovakia,}
\centerline{\small e-mail:jenca@mat.savba.sk}
\vskip 1cm

\noindent
{\bf Abstract.}{\small We present a construction of a Banach manifold on the set of
faithful normal states of a von Neumann algebra, where the underlying Banach space is
 a quantum analogue of an Orlicz space. On the manifold, we introduce the exponential and mixture connections
 as dual pair of affine connections. }

\section{Introduction}

An information manifold is a family of states of some
classical or quantum system, endowed with a differentiable manifold structure.
For  finitely parametrized families of probability distributions,
the geometry of such manifolds and its applications in parameter estimation
is already well understood, see for example the books
\cite{amana,abklr}.

The non-parametric version was introduced by Pistone and Sempi
\cite{pise,gipi}, based on the theory of Orlicz spaces. For the
quantum version, some proposals for infinite dimensional manifold
structure can be found in \cite{grastre,streater00a,streater00b}.

The aim of this paper is to introduce a differentiable manifold structure
on the set of faithful states of a quantum system, represented by a von Neumann
algebra $\iM$. Moreover, we want this manifold to be a quantum counterpart of the
Pistone and Sempi construction.

We use an approach similar to Grasselli \cite{grasselli05} in the
commutative case: we define an Orlicz norm on the space of
self-adjoint operators in $\iM$ and take the completion under this
norm to be the underlying Banach space for the manifold. The norm
is defined by a quantum Young function, as in
\cite{streater04}. The definition of a Young function on a Banach
space, together with some results on the associated norms, can be
found in Section \ref{sec:young}. For a faithful state $\vfi$, the
quantum Orlicz space $B_\vfi$ and its centered version
$B_{\vfi,0}$ are introduced in Section \ref{sec:orlicz}. The
definition of the related Young function is based on the relative
entropy approach to state perturbation. We treat the dual spaces
in Section \ref{sec:dual}.
 The main result is contained in Section
\ref{sec:mani}, where the manifold structure is introduced and,
moreover, the exponential and mixture connections are defined as a
pair of dual affine connections on each  connected component of
the manifold.

\section{Preliminaries.}

We recall some properties of relative entropy and perturbed
states, that will be needed later. See \cite{ohypetz} for details.

Let $\iM$ be a von Neumann algebra in standard form. For $\omega$
and $\vfi$ in $\iM_*^+$, the relative entropy is defined as
$$
S(\omega,\vfi)=\left\{\begin{array}{lc} -\langle\log
(\Delta_{\vfi,\xi_\omega})\xi_\omega,\xi_\omega\rangle & \ \mbox{if } \supp
\omega\le\supp \vfi\\ \infty & \mbox{otherwise}
\end{array}\right.
$$
where $\xi_\omega$ is the representing vector of $\omega$ in a natural positive
cone and $\Delta_{\vfi,\xi_\omega}$ is the relative modular operator. Then $S$
is jointly convex and weakly lower semicontinuous.
Let us denote $\iP_\vfi:=\{ \omega\in\iM_*^+, S(\omega,\vfi)<\infty\}$, then
$\iP_\vfi$ is a convex cone. We will need the following Donald's identity
\begin{equation}\label{eq:donald}
S(\psi,\vfi)+
\sum_iS(\psi_i,\psi) =\sum_iS(\psi_i,\vfi)
\end{equation}
where $\psi_i\in \iM_*^+$, $i=1,\dots,n$, and
$\psi=\sum_i\psi_i$. Since $S(\psi_i,\psi)$ is always finite,
it follows from this identity that $\sum_i\psi_i\in
\iP_\vfi$ if and only if $\psi_i\in \iP_\vfi$ for all $i$.

Let $\mathfrak{S}_*$
be the set of normal states on $\iM$
and let $\iS_\vfi:=\{\omega\in \mathfrak{S}_*,\ S(\omega,\vfi)<\infty\}$. Then
$\iS_\vfi$ is a convex set and
generates $\iP_\vfi$. From (\ref{eq:donald}), we get
\begin{equation}\label{eq:donalds}
S(\psi_\lambda,\vfi)+\lambda
S(\psi_1,\psi_\lambda)+(1-\lambda)S(\psi_2,\psi_\lambda)
 =\lambda S(\psi_1,\vfi)+(1-\lambda)S(\psi_2,\vfi)
\end{equation}
where $\psi_1$, $\psi_2$ are normal states and
$\psi_\lambda=\lambda\psi_1+(1-\lambda)\psi_2$, $0\le \lambda\le 1$.
As above, it  follows that $\psi_\lambda\in \iS_\vfi$ if and
only if both $\psi_1,\psi_2\in \iS_\vfi$,  in other words, $\iS_\vfi$ is a face
in $\mathfrak{S}_*$. For $C>0$, we define the set $\iS_C:=\{\omega,\
S(\omega,\vfi)\le C\}$. Then $\iS_C$ is convex and  compact in the
$\sigma(\iM_*,\iM)$ topology.

Let us suppose that $\vfi$ is a faithful normal state on $\iM$ and let $h$ be a
self-adjoint element in $\iM$.
The perturbed state $[\vfi^h]$  is defined as the unique maximizer  of
\begin{equation}\label{eq:sup}
\sup_{\omega\in \mathfrak{S}_*}\{ \omega(h)-S(\omega,\vfi)\}
\end{equation}
Then $[\vfi^h]$ is a faithful normal state and $S([\vfi^h],\vfi)$ is finite.
Let $c_\vfi(h)$ be the  supremum in (\ref{eq:sup}), that is
\begin{equation} \label{eq:cfi}
c_\vfi(h)=[\vfi^h](h)-S([\vfi^h],\vfi)
\end{equation}
It is known that
\begin{equation}\label{eq:gt}
\vfi(h)\le c_\vfi(h)\le \log \vfi(e^h)
\end{equation}
Moreover, we have
\begin{equation}\label{eq:sup2}
\omega(h)-S(\omega,\vfi)=c_\vfi(h)-S(\omega,[\vfi^{h}])
\end{equation}
for any self-adjoint $h\in\iM$ and $\omega\in\mathfrak{S}_*$.
Let $h,k$ be self-adjoint elements in $\iM$, then the chain rule $[\vfi^{h+k}]=[[\vfi^h]^k]$ and
\begin{equation} \label{eq:chainM}
c_\vfi(h+k)=c_{[\vfi^h]}(k)+c_\vfi(h)
\end{equation}
holds. Let now $\xi_\vfi$ be the vector representative of $\vfi$ and let
$\vfi^h\in\iM_*^+$ be
the functional induced by the perturbed vector
$$
\xi_\vfi^h:=e^{\frac12(\log\Delta_{\vfi}+h)}\xi_\vfi=
e^{c_\vfi(h)}\Delta^{1/2}_{[\vfi^h],\vfi}\xi_\vfi
$$
Then  $c_\vfi(h)=\log \vfi^h(1)$ and $[\vfi^h]=\vfi^h/\vfi^h(1)$.
Moreover, if $\vfi^h=\vfi^k$, then $h=k$.

\section{Young functions on Banach spaces and the associated norms.}\label{sec:young}

Let $V$ be a real Banach space and let $V^*$ be its dual, with the duality
pairing $\langle v,x\rangle= v(x)$. Recall that any convex lower semicontinuous
function $V\to \mathbb R\cup\{+\infty\}$ is lower semicontinuous with respect
to the $\sigma(V,V^*)$-topology.

\subsection*{The Young function.}
 We will say that a function
$\Phi:V\to \mathbb{R}\cup \{\infty\}$ is a Young function, if it satisfies:
\begin{enumerate}
\item[(i)] $\Phi$ is convex and lower semicontinuous,
\item[(ii)] $\Phi(x)\ge 0$ for all $x\in V$ and $\Phi(0)=0$
\item[(iii)] $\Phi(x)=\Phi(-x)$ for all $x\in V$
\item[(iv)] if $x\ne 0$, then $\lim_{t\to\infty}\Phi(tx)=\infty$
\end{enumerate}

\begin{lemma}\label{lemma:cramer} Let $\Phi$ be a Young function.
Let us define the sets
\begin{eqnarray*}
C_\Phi&:=&\{ x\in V, \Phi(x)\le 1\}\\
L_\Phi&:=&\{ x\in V, \exists s>0, \mbox{ such that  } \Phi(sx)<\infty\}.
\end{eqnarray*}
Then $C_\Phi$ is absolutely convex and $L_\Phi=\cup_nnC_\Phi$. In particular,
$L_\Phi$ is a (real) vector space.
\end{lemma}

{\it Proof.} Let $x,y\in C_\Phi$ and let  $\alpha,\beta\in \mathbb{R}$, such
that $|\alpha|+|\beta|\le 1$. Put $\gamma=1-|\alpha|-|\beta|$, then
$$
\Phi(\alpha x+\beta y)=\Phi(|\alpha|{\rm sgn}(\alpha)x+|\beta|{\rm sgn}(\beta)y
+\gamma 0)\le |\alpha|\Phi(x)+|\beta|\Phi(y)\le 1
$$
hence $\alpha x+\beta y\in C_\Phi$ and $C_\Phi$ is absolutely convex.

Let now $x\in L_\Phi$ and let $s>0$ be such that $\Phi(sx)=K<\infty$. Choose
$m\in \mathbb{N}$ such that $m\ge \max\{1/s,K/s\}$, then by convexity
$$
\Phi(\frac 1m x)=\Phi(\frac1{ms}sx)\le \frac1{ms}\Phi(sx)=\frac K{ms}\le 1
$$
and $x\in mC_\Phi$. Since obviously $nC_\Phi\subset L_\Phi$ for all $n$, we have
$L_\Phi=\cup_nnC_\Phi$, which clearly implies that $L_\Phi$ is a vector space.
\qed

Let us recall that the effective domain
$$
{\rm dom} (\Phi)=\{ x\in V,\ \Phi(x)<\infty\}
$$
is a convex set. Any convex lower semicontinuous function is continuous
in the interior of its effective domain, \cite{ektem}.   Clearly,
$L_\Phi$ is the smallest vector space containing ${\rm dom}(\Phi)$.

In the space $L_\Phi$, we now introduce the Minkowski functional of $C_\Phi$,
$$
\|x\|_\Phi:=\inf\{\rho>0, x\in \rho C_\Phi\}.
$$
Since $C_\Phi$ is absolutely
convex and absorbing, $\|\cdot\|_\Phi$ is a seminorm. Moreover,
$\|x\|_\Phi=0$ means that $\Phi(tx)\le 1$ for all $t>0$. By the property (iv),
this implies that $x=0$. It follows that
$\|\cdot\|_\Phi$ defines a norm in $L_\Phi$. Let us denote by $B_\Phi$ the
completion of $L_\Phi$ under this norm.

\begin{lemma}\label{lemma:Phinorm}
Let $x\in L_\Phi$. Then $\|x\|_\Phi\le 1$ if and only if  $\Phi(x)\le 1$.
\end{lemma}

{\it Proof.}
If  $\|x\|_\Phi<1$,  then $x\in C_\Phi$ and $\Phi(x)\le 1$. Let
now  $\|x\|_\Phi=1$ and let $t_n< 1$ be a sequence converging to 1. Then
$\Phi(t_nx)\le 1$ for all $n$ and, by lower semicontinuity,
$\Phi(x)\le \liminf_n \Phi(t_nx)\le 1$. Hence
$\|x\|_\Phi\le 1$ implies $\Phi(x)\le 1$. On the other hand, if $\Phi(x)\le 1$,
then $x\in C_\Phi$ and clearly $\|x\|_\Phi\le 1$.
\qed

\begin{lemma}\label{lemma:normandPhi} Let $x\in L_\Phi$. Then
$\|x\|_\Phi\le 1$ implies  $\Phi(x)\le \|x\|_\Phi$ and
$\|x\|_\Phi> 1$  implies $ \Phi(x)\ge \|x\|_\Phi$. Moreover, if $\Phi$ is finite
valued, then $\|x\|_\Phi=1$ if and only if $\Phi(x)=1$.
\end{lemma}

{\it Proof.} Let $\|x\|_\Phi\le 1$. By convexity of $\Phi$ and Lemma
\ref{lemma:Phinorm},
$$
\Phi(x)=\Phi(\|x\|_\Phi\frac x{\|x\|_\Phi})\le \|x\|_\Phi\Phi(\frac
x{\|x\|_\Phi})\le\|x\|_\Phi
$$

Let now $\|x\|_\Phi>1$, then $\Phi(x)>1$.
If $\Phi(x)=\infty$, then the assertion is obviously
true. Let us suppose that $\Phi(x)$ is finite. The function
$t\mapsto\Phi(tx)$ is  convex and bounded on $<0,1>$, hence continuous on
$(0,1)$. It follows that $\Phi(tx)=1$ for some  $t$ in this
interval and clearly $t=\frac 1{\|x\|_\Phi}$. We have
$$
1=\Phi(tx)\le t\Phi(x)
$$
and hence $\|x\|_\Phi\le \Phi(x)$.
This also proves that last  statement. \qed

\subsection*{The conjugate function.} Let $V^*$ be the dual space.
Let the function $\Phi^*:V^*\to
\mathbb{R}\cup\{\infty\}$ be the conjugate of $\Phi$,
$$
\Phi^*(v)=\sup_{x\in V}\{v(x)-\Phi(x)\}=\sup_{x\in {\rm Dom}(\Phi)}
\{v(x)-\Phi(x)\}
$$
The function $\Phi^*$ is convex, lower semicontinuous and positive,
$\Phi^*(v)=\Phi^*(-v)$ and $\Phi^*(0)=0$. But, in general, $\Phi^*$ is not a Young
function: consider the case when $\Phi(0)=0$ and $\Phi(x)=\infty$ for all $x\ne
0$, then $\Phi$ is a Young function, but its conjugate is identically equal to 0
on $V^*$ and the condition (iv) is not satisfied.

Let $({\rm dom}(\Phi))^\perp$ be the orthogonal subspace to ${\rm dom}(\Phi)$ in
$V^*$, that is
$$
({\rm dom}(\Phi))^\perp:=\{ v\in V^*,\ v(x)=0 \ \mbox{for all } x\in {\rm
dom}(\Phi)\}
$$
Then $({\rm dom}(\Phi))^\perp$ is a closed subspace in $V^*$. Let $\tilde{V}$ be
the quotient space $\tilde V=V^*/_{({\rm dom}(\Phi))^\perp}$. If $u$ and $v$ are
elements in the same equivalence class, then
$$
\Phi^*(v)=\sup_{x\in {\rm dom}(\Phi)}\{v(x)-\Phi(x)\}=\sup_{x\in {\rm
dom}(\Phi)}\{u(x)-\Phi(x)\}=\Phi^*(u)
$$
and $\Phi^*$ is well defined as a function on $\tilde V$.

\begin{lemma}  $\Phi^*: \tilde V\to \mathbb{R}\cup\{\infty\}$ is a
Young function.
\end{lemma}

{\it Proof.}
It is easy to see that $\Phi^*$
 satisfies (i), (ii) and (iii) from
the definition of a Young function.
Moreover, it follows from the definition of the conjugate function
that
\begin{equation}\label{eq:young}
|v(x)|\le \Phi(x)+\Phi^*(v),\quad \mbox{for all } x\in V, v\in \tilde V
\end{equation}
Let $v\in \tilde V$, $v\ne 0$. Then there is an element  $x\in {\rm  dom}(\Phi) $
such that $v(x)\ne 0$.
It follows that $\Phi^*(tv)\geq |tv(x)|-\Phi(x)$ for all $t$  and (iv) is satisfied.
\qed

We will  define  $C_{\Phi^*}$,  $L_{\Phi^*}$, $\|\cdot\|_{\Phi^*}$ and
$B_{\Phi^*}$
in the same way as for $\Phi$.

\begin{lemma} (H\"older inequality).
$$
|v(x)|\le 2\|x\|_\Phi\|v\|_{\Phi^*}\qquad \mbox{for all }\ x\in B_\Phi, v\in B_{\Phi^*}
$$
\end{lemma}

{\it Proof.} Let $x\in C_\Phi$, $v\in C_{\Phi^*}$, then by (\ref{eq:young})
$$
|v(x)|\le \Phi(x)+\Phi^*(v)\le 2
$$
Let $x\in L_\Phi$, $v\in L_{\Phi^*}$. By Lemma \ref{lemma:Phinorm},
$\frac x{\|x\|_\Phi}\in C_\Phi$,
$\frac v{\|v\|_{\Phi^*}}\in C_{\Phi^*}$ and therefore
$|v(x)|\le 2\|x\|_\Phi\|v\|_{\Phi^*}$. Clearly, the inequality extends to
$x\in B_\Phi$, $v\in B_{\Phi^*}$.
\qed

\subsection*{The second conjugate.} If $E$ is a Banach space and $H\subset E$ is
a closed subspace, then the dual of the quotient space $(E/H)$ can be identified with
$H^\perp$. It follows that $\tilde V^*\cap V=({\rm dom}(\Phi))^{\perp\perp}$,
which is nothing else than the closure of $L_\Phi$ in $V$. Let us denote this
space by $\bar V$.

As before, we can find the conjugate function to $\Phi^*:\tilde V\to \mathbb
R\cup\{+\infty\}$ with respect to the pair $(\tilde V,\tilde V^*)$.
Note that for $x$ in $\bar V$, we have
$$
\sup_{v\in \tilde V}\{v(x)-\Phi^*(v)\}=\sup_{v\in V^*}\{v(x)-\Phi^*(v)\}=
\Phi^{**}(x)
$$
where  $\Phi^{**}$ is the  second conjugate to $\Phi:V\to \mathbb
R\cup\{+\infty\}$. Since  $\Phi$ is
convex and lower semicontinuous, $\Phi^{**}(x)=\Phi(x)$ on $V$, \cite{ektem}.
It follows in particular that the restriction of $\Phi^{**}$ to $\bar V$ is a
Young function.

It is clear from H\"older inequality that any $x\in L_\Phi$
defines a bounded linear functional on $B_{\Phi^*}$.
Let $\|x\|^*_{\Phi^*}$ be its norm in $B_{\Phi^*}^*$, then by  Lemma
\ref{lemma:Phinorm},
$$\|x\|^*_{\Phi^*}=\sup \{|v(x)|, \Phi^*(v)\le 1\}. $$
Similarly, if $v\in L_{\Phi^*}$, then $v\in B^*_\Phi$ and we have
$$
\|v\|^*_\Phi=\sup \{|v(x)|,
\Phi(x)\le 1\}
$$

\begin{lemma}\label{lemma:equivalent} For $x\in L_\Phi$, we have
$\|x\|_\Phi\le \|x\|_{\Phi^*}^*\le 2 \|x\|_\Phi$. Similarly, if
$v\in L_{\Phi^*}$,
then $\|v\|_{\Phi^*}\le \|v\|^*_\Phi\le 2\|v\|_{\Phi^*}$.
\end{lemma}

{\it Proof.} Let $v\in L_{\Phi^*}$. By H\"older inequality,
$\|v\|_\Phi^*\le 2\|v\|_{\Phi^*}$. Let now $\|v\|^*_\Phi=1$, then
for $x\in C_\Phi$ we have
$$v(x)-\Phi(x)\le 1.$$
On the other hand, for $x\in {\rm dom}(\Phi)$, such that $\Phi(x)>1$,
we get from Lemma \ref{lemma:normandPhi}
$$v(x)-\Phi(x)\le v(x)-\|x\|_\Phi\le0.$$
It follows that
${\Phi^*}(v)\le 1$
and $v\in C_{\Phi^*}$, hence $\|v\|_{\Phi^*}\le 1$. Therefore,
$\|v\|_{\Phi^*}\le\|v\|^*_\Phi$ for all $v\in L_{\Phi^*}$. The proof for
$x\in L_\Phi$ is the same, using the fact that $\Phi$ is the conjugate of
$\Phi^*$.
\qed

\begin{prop}\label{prop:dual} $B_{\Phi^*}\subseteq B^*_\Phi$ and
 $L_{\Phi^*}=\tilde V\cap B_\Phi^*$.
Similarly, $B_\Phi\subseteq B^*_{\Phi^*}$ and $L_\Phi=\bar V\cap B^*_{\Phi^*}$.
\end{prop}

{\it Proof.} As we have seen, $L_{\Phi^*}$ is a vector subspace in $B_\Phi^*$
and the norms
in $L_{\Phi^*}$ and $B_\Phi^*$ are equivalent, hence $B_{\Phi^*}\subseteq B^*_\Phi$.
Let now  $v\in \tilde V\cap B^*_\Phi$ be such that $\|v\|^*_\Phi=1$.
Then  ${\Phi^*}(v)\le 1$, exactly as in the proof of Lemma \ref{lemma:equivalent}.
It follows that for all  $v\in \tilde V\cap B^*_\Phi$,
${\Phi^*}(v/\|v\|^*_\Phi)\le1<\infty$
and $v\in L_{\Phi^*}$. Again, the proof for $L_\Phi$ and $B_\Phi$ is the same.
\qed

Let $\Phi$ be a Young  function such that 0 is an
interior point in  ${\rm dom}(\Phi)$.  Then the function $\Phi$ is continuous
in 0, therefore there is an open set $U$ containing 0  such that
$U\subset C_\Phi$. It follows
that $C_\Phi$ is a neighborhood of $0$ in $V$, hence it is absorbing in $V$
\begin{equation}\label{eq:Velfi}
V=\cup_nnC_\Phi=L_\Phi\quad \mbox{ (as sets)}
\end{equation}
Since $C_\Phi$ is a convex body (that is, 0 is a topological interior point),
its Minkowski functional $\|\cdot\|_\Phi$ is continuous with respect to the
original norm (\cite{kothe}, p. 182). It follows that
we have the continuous inclusion $V\sqsubseteq B_\Phi$. Further,
since ${\rm dom}(\Phi)$ has non-empty interior, $({\rm dom}(\Phi))^\perp=\{0\}$
and $\tilde V=V^*$. Clearly also $\bar V=V$.

\begin{prop}\label{prop:interior} Let  $0\in {\rm int}\, {\rm dom}(\Phi)$. Then
$V\sqsubseteq B_\Phi\subseteq B_{\Phi^*}^*$ and
$L_{\Phi^*}=B_{\Phi^*}=B_\Phi^*\sqsubseteq V^*$.
\end{prop}

{\it Proof.}
By (\ref{eq:Velfi}), each $x\in V$ is in $L_\Phi$, and by continuity,
 $\|x\|_\Phi\le K\|x\|$, for some fixed $K>0$.
Let $v\in B^*_\Phi$, then
$$
|v(x)|\le \|v\|_\Phi^*\|x\|_\Phi\le K\|v\|_\Phi^*\|x\| \quad \mbox{for }\ x\in V
$$
hence $v\in V^*=\tilde V$ and $\|v\|^*\le K\|v\|_\Phi^*$.
The statement now  follows from  Proposition \ref{prop:dual}.
\qed

\section{The spaces $B_\vfi$ and $B_{\vfi,0}$.}\label{sec:orlicz}

Let $\iM_s$ be the real Banach subspace of self-adjoint elements in $\iM$, then
the dual $\iM_s^*$ is the subspace of hermitian (not necessarily normal)
functionals in $\iM^*$. We define the functional $F_\vfi:\ \iM_s^*\to
\mathbb{R}\cup\{\infty\}$ by
$$
F_\vfi(\omega)=\left\{\begin{array}{lc} S(\omega,\vfi) & \ \mbox{if
}\omega\in\mathfrak{S}_* \\ \infty & \mbox{otherwise}
\end{array}\right.
$$
Then $F_\vfi$ is convex and  lower semicontinuous, with ${\rm dom
}(F_\vfi)=\iS_\vfi$. It follows from (\ref{eq:donald}) that $F_\vfi$ is strictly
convex. Its conjugate $F^*_\vfi$ is
$$
F^*_\vfi(h)=\sup_{\omega\in
\mathfrak{S}_*}\{\omega(h)-F_\vfi(\omega)\}=c_\vfi(h), \quad h\in\iM_s
$$
Hence $c_\vfi$ is convex and lower semicontinuous, in fact, since
finite valued, it is continuous on $\iM_s$. We have
$c^*_\vfi=F^{**}_\vfi=F_\vfi$ on $\iM_s^*$. Note also that
\begin{equation}\label{eq:cadit}
c_\vfi(h+\lambda)=c_\vfi(h)+\lambda,\qquad \forall \lambda\in\mathbb{R}
\end{equation}

We define another convex and lower semicontinuous functional on $\iM_s^*$,
namely,
$$
\bar F_\vfi(\omega)=\left\{\begin{array}{lc} S(\omega,\vfi)-\omega(1) & \ \mbox{if
}\omega\in\iM^+_* \\ \infty & \mbox{otherwise}
\end{array}\right.
$$
Then the conjugate functional is
\begin{eqnarray*}
\bar F_\vfi^*(h)&=&\sup_{\omega\in
\iM^+_*}\{\omega(h)-S(\omega,\vfi)+\omega(1)\}=
\sup_{\omega\in \mathfrak{S}_*,\lambda\in \mathbb{R}^+}
\{\lambda\omega(h)-S(\lambda\omega,\vfi)+\lambda\}=\\
&=&\sup_{\omega\in \mathfrak{S}_*,\lambda\in \mathbb{R}^+}
\{\lambda(\omega(h)-S(\omega,\vfi))-\lambda\log\lambda+\lambda\}=\\
&=&\sup_{\lambda\in\mathbb{R}^+}\{\lambda(c_\vfi(h)+1)-\lambda\log\lambda\}=e^{c_\vfi(h)}
=\vfi^h(1)
\end{eqnarray*}
Again, $h\mapsto\vfi^h(1)$ is convex and continuous and $\bar
F_\vfi^{**}=\bar F_\vfi$. Next, we define a Young
function on $\iM_s$.

Let $\Phi_\vfi:\ \iM_s\to \mathbb{R}^+$ be defined by
$$
\Phi_\vfi(h)=\frac {\vfi^h(1)+\vfi^{-h}(1)}2-1
$$

\begin{lemma}\label{lemma:Phi}
$\Phi_\vfi$ is a Young function.
\end{lemma}

{\it Proof.} The property (i) from the definition of a Young function follows
from the properties of $h\mapsto \vfi^h(1)$. Since
$\vfi^h(1)=e^{c_\vfi(h)}\geq e^{\omega(h)-S(\omega,\vfi)}$ for all normal states
$\omega$,  we have
\begin{equation}\label{eq:helpful}
\Phi_\vfi(h)\geq \cosh(\omega(h))e^{-S(\omega,\vfi)}-1
\end{equation}
In particular,
\begin{equation}\label{eq:vfi}
\Phi_\vfi(h)\ge \cosh(\vfi(h))-1\ge 0\quad \mbox {for all }\ h
\end{equation}
Since obviously $\Phi_\vfi(0)=0$, (ii) follows.
Let now $h$ be such that $\omega(h)=0$ for all $\omega\in \iS_\vfi$, then
by definition, $c_\vfi(h)=0$ and $\vfi=\vfi^h$, hence $h=0$.
Therefore if $h\ne 0$, then there is a state $\omega\in\iS_\vfi$
such that $\omega(h)\ne 0$ and then
$\lim_{t\to\infty}\cosh(t\omega(h))=\infty$, this implies (iv).
Property (iii) is obviously satisfied.
\qed

Let $C_\vfi:=C_{\Phi_\vfi}$, $B_\vfi:=B_{\Phi_\vfi}$ and
$\|\cdot\|_\vfi:=\|\cdot\|_{\Phi_\vfi}$. Since ${\rm dom}\Phi_\vfi=\iM_s$,
we have by Proposition \ref{prop:interior} that
$\iM_s\sqsubseteq B_\vfi$. If $\Phi_\vfi^*$ is the conjugate of $\Phi_\vfi$,
then
$B^*_\vfi=B_{\Phi_\vfi^*}\sqsubseteq \iM^*_s$.

Let now $h\in \iM_s$, such that $\|h\|_\vfi=t>0$, that is,
$$
\Phi_\vfi(\frac{h}t)=1
$$
If $\omega$ is a state,  then  by (\ref{eq:helpful}),
\begin{equation}\label{eq:omega}
\cosh(\frac{\omega(h)}t)\le 2e^{S(\omega,\vfi)}
\end{equation}
If  $\omega\in \iS_\vfi$,  then $|\omega(h)|\le ct$, where $c>0$ is
some constant depending  on $S(\omega,\vfi)$. It follows that
each $\omega\in \iS_\vfi$  extends to a  continuous linear functional
on $B_\vfi$. Moreover,
 for $C>0$,  $\iS_C$ is an  equicontinuous subset in $B_\vfi^*$.

Let $\iM_{s,0}\subset \iM_s$ be the subspace of  elements satisfying
$\vfi(h)=0$.  Then by putting $\omega=\vfi$ in (\ref{eq:sup2}), we get
$$
c_\vfi(h)=S(\vfi,[\vfi^h])\geq0
$$
Let us define
$$
\Phi_{\vfi,0}(h)=\frac{c_\vfi(h)+c_\vfi(-h)}2,\quad h\in \iM_{\vfi,0}
$$
Then  it is easy to check that $\Phi_{\vfi,0}$ is a  Young
function on $\iM_{\vfi,0}$.  We have
\begin{lemma}\label{lemma:finula} Let $h\in \iM_{s,0}$. Then
$$
\Phi_{\vfi,0}(h)\le \Phi_\vfi(h)\le e^{2\Phi_{\vfi,0}}-1
$$

\end{lemma}

{\it Proof.}
The first inequality follows from $a\le e^a-1$ for $a\ge 0$, the
second follows from $x+y\le 2xy$ for $x,y\ge1$.
\qed

Let us construct the Banach space $B_{\Phi_{\vfi,0}}=:B_{\vfi,0}$ and let
$\|\cdot\|_{\vfi,0}:=\|\cdot\|_{\Phi_{\vfi,0}}$.

\begin{prop}\label{prop:finula} The norms $\|\cdot\|_{\vfi,0}$ and
$\|\cdot\|_\vfi$
are equivalent on $\iM_{s,0}$.
\end{prop}

{\it Proof.}
Let us denote $C_{\vfi,0}:=C_{\Phi_{\vfi,0}}$. We show that
\begin{equation} \label{eq:cf0}
\frac12{\log2}\,C_{\vfi,0}\subseteq C_\vfi\cap \iM_{s,0}\subseteq C_{\vfi,0}
\end{equation}
 Let $h\in C_{\vfi,0}$ and  $t=\frac12\log2$. Then by convexity, $\Phi_{\vfi,0}(th)\le t=\frac12\log2$
and hence
$$\Phi_\vfi(th)\le e^{2\Phi_{\vfi,0}(th)}-1\le 1,$$
which implies $tC_{\vfi,0}\subseteq C_\vfi\cap \iM_{s,o}$. The other inclusion follows from
the first inequality in Lemma \ref{lemma:finula}.
It follows from (\ref{eq:cf0}) that for $h\in \iM_{s,0}$,
$$
 \|h\|_{\vfi,0}\le \|h\|_{\vfi}\le \frac2{\log2}\|h\|_{\vfi,0}
$$

\qed

Note that since $\vfi\in\iS_\vfi$, $\vfi$ extends to a bounded linear functional
on $B_\vfi$. Clearly, the completion of
$\iM_{s,0}$ under the norm $\|\cdot\|_\vfi$ is the Banach subspace $\{h\in
B_\vfi,\ \vfi(h)=0\}$. It follows from the above Proposition that $B_{\vfi,0}$
can be identified with the subspace of centered elements in $B_\vfi$.

\section{Extension of   $c_\vfi$.}

Since $\iS_\vfi\subset B^*_\vfi\sqsubseteq \iM^*_s$, the restriction of
$F_\vfi$ is a strictly convex lower semicontinuous functional on $B_\vfi^*$,
with effective domain $S_\vfi$. Its conjugate $F^*_\vfi$ is a lower
semicontinuous extension of $c_\vfi$ to $B_\vfi$, moreover,
$F^{**}_\vfi=F_\vfi$.
We show that this extension has again
values in $\mathbb R$ and is continuous.

\begin{lemma}\label{lemma:bounded}  Let the sequence
$\{h_n\}_n\subset \iM_s$ be Cauchy
in the norm $\|\cdot\|_\vfi$.
Then the sequences $\{c_\vfi(h_n)\}_n$ and  $\{S([\vfi^{h_n}],\vfi)\}_n$
are bounded.
\end{lemma}

{\it Proof.}  By  (\ref{eq:gt}), we have for all $n$
$$
\vfi(h_n)\le c_\vfi(h_n)
$$
Since $\vfi(h_n)$ converges, $c_\vfi(h_n)$ is bounded from below. Further,
let $n_0$ be such that $\|h_n-h_{n_0}\|_\vfi<1$ for all $n\ge n_0$, then
$$
\omega(h_n)-S(\omega,\vfi)\le\omega(h_{n_0})+c_\vfi(h_n-h_{n_0})\le \|h_{n_0}\|+
\log 2
$$
for all such $n$ and  $\omega\in S_\vfi$.
It follows that $\{c_\vfi(h_n)\}_n$ is bounded.

If $\{h_n\}_n$ is Cauchy, then the sequence $\{th_n\}_n$ is also Cauchy
for all $t\in \mathbb{R}$ and
there are  constants $A_t, B_t$,  such that
$$
A_t\le c_\vfi(th_n)\le B_t,\qquad \forall n
$$
On the other hand , we have
$$
\frac d{dt}c_\vfi(th_n)|_{t=1}=[\vfi^{h_n}](h_n)
$$
By convexity,
$$
c_\vfi(th_n)\ge c_\vfi(h_n)+(t-1)\frac d{dt} c_\vfi(th_n)|_{t=1}\ge A_1+
(t-1)[\vfi^{h_n}](h_n)
$$
For arbitrary fixed $t>1$, we get
$$
[\vfi^{h_n}](h_n)\le\frac{B_t-A_1}{t-1},\qquad \forall n
$$
Boundedness of $S([\vfi^{h_n}],\vfi)$ now follows from
$$
0\le S([\vfi^{h_n}],\vfi)=[\vfi^{h_n}](h_n)-c_\vfi(h_n).
$$
\qed

\begin{thm}\label{thm:perturb} Let  $\{h_n\}_n$ be a sequence in $\iM_s$,
converging to some $h$ in $B_\vfi$. Then
\begin{equation}\label{eq:maximizer}
\lim_n c_\vfi(h_n)=\sup_{\omega\in \iS_\vfi}\{\omega(h)-S(\omega,\vfi)\}
\end{equation}
and there  is a unique state $\psi\in \iS_\vfi$ such that
the supremum is attained.  The state $\psi$ is faithful.
Moreover,
$\lim_n S([\vfi^{h_n}],\vfi)=S(\psi,\vfi)$, $\lim_n [\vfi^{h_n}(h_n)]=\psi(h)$
and $\lim_n S(\psi,[\vfi^{h_n}])=0$. In particular, $[\vfi^{h_n}]$
converges to $\psi$ in norm.
\end{thm}

The state $\psi$ will be denoted by $[\vfi^h]$ and
the limit $\lim_nc_\vfi(h_n)=:c_\vfi(h)$.

{\it Proof.} This proof is similar to the proof of Theorem 12.3. in \cite{ohypetz}.

By Lemma \ref{lemma:bounded}, there is some $C>0$ such that
$[\vfi^{h_n}]\in\iS_{\vfi,C}$ for all $n$.
The set $\iS_{\vfi,C}$ is weakly relatively compact in $\mathfrak{S}_*$ and
hence
there is  subsequence $[\vfi^{h_{n_k}}]$  converging weakly to a state
$\psi\in \iS_{\vfi,C}$. We will show that $[\vfi^{h_{n_k}}](h_{n_k})$ converges
to $\psi(h)$.

Since $\iS_C$ is an equicontinuous subset in $B_\vfi^*$,
$\omega(h_n)$ converges to $\omega(h)$ uniformly for all  $\omega\in\iS_{\vfi,C}$.
This implies
$$
|[\vfi^{h_{n_k}}](h_{n_k})-[\vfi^{h_{n_k}}](h)|<\varepsilon
$$
for sufficiently large $n_k$.
We further have
\begin{eqnarray*}
|[\vfi^{h_{n_k}}](h)-\psi(h)|&\le&|[\vfi^{h_{n_k}}](h)-[\vfi^{h_{n_k}}](h_m)|\\
&+&
|[\vfi^{h_{n_k}}](h_m)-\psi(h_m)|+|\psi(h_m)-\psi(h)|<\varepsilon
\end{eqnarray*}
for  sufficiently large  $m$ and $n_k$. Putting both inequalities
together, we get
 $[\vfi^{h_{n_k}}](h_{n_k})\to \psi(h)$.

Let $\omega\in \iS_\vfi$. By definition,
$$
[\vfi^{h_{n_k}}](h_{n_k})-S([\vfi^{h_{n_k}}],\vfi)=c_\vfi(h_{n_k})\ge
\omega(h_{n_k})-S(\omega,\vfi)
$$
By weak lower semicontinuity of the relative entropy, we get
\begin{equation}\label{eq:limsup}
\psi(h)-S(\psi,\vfi)\ge \limsup c_\vfi(h_{n_k})\ge\omega(h)-S(\omega,\vfi)
\end{equation}
and thus $\psi$ is a maximizer of (\ref{eq:maximizer}). On the other hand,
$$
\psi(h_{n_k})-S(\psi,\vfi)\le [\vfi^{h_{n_k}}](h_{n_k})-
S([\vfi^{h_{n_k}}],\vfi)=c_\vfi(h_{n_k})
$$
From this and (\ref{eq:limsup}), it follows that
$\psi(h)-S(\psi,\vfi)=\lim c_\vfi(h_{n_k})$.
It also follows that
$$
\limsup S([\vfi^{h_{n_k}}],\vfi)\le S(\psi,\vfi)
$$
and this, together with lower semicontinuity implies that
$S([\vfi^{h_{n_k}}],\vfi)$ converges to $S(\psi,\vfi)$.

To show that such $\psi$ is unique, suppose that $\psi'$ is another maximizer,
then for $\psi_\lambda:=\lambda\psi+(1-\lambda)\psi'$,
$0\le\lambda\le 1$, we have
$$
\psi(h)-S(\psi,\vfi)\ge\psi_\lambda(h)-S(\psi_\lambda,\vfi)\ge
\psi_\lambda(h)-\lambda
S(\psi,\vfi)-(1-\lambda)S(\psi',\vfi)=\psi(h)-S(\psi,\vfi)
$$
hence $\psi_\lambda$ is a maximizer as well and, moreover,
$$
S(\psi_\lambda,\vfi)=\lambda S(\psi,\vfi)+(1-\lambda)S(\psi',\vfi)
$$
By strict convexity, his implies that $\psi=\psi'$. It also
follows that the whole sequence $[\vfi^{h_n}]$ converges weakly to
$\psi$.

Using (\ref{eq:sup2}), we have
$$
S(\vfi,\psi)\le \liminf_n S(\vfi,[\vfi^{h_n}])=\lim_n c_\vfi(h_n)-\vfi(h)<\infty
$$
This implies that $\supp\vfi\le\supp \psi$ and $\psi$ is faithful.
Finally, by taking the limit in the equality,
$$
\psi(h_n)-S(\psi,\vfi)=c_\vfi(h_n)-S(\psi,[\vfi^{h_n}])
$$
we get $\lim_nS(\psi,[\vfi^{h_n}])\to 0$.\qed

\begin{coro} Let $h_n$ be a sequence in $B_\vfi$, then $h_n\to 0$
if and only if $c_\vfi(th_n)\to 0$ for all $t\in \mathbb{R}$.
\end{coro}

{\it Proof.} Let $h_n$ be such that
$c_\vfi(th_n)=\log \vfi^{th_n}(1)$ converges to 0,
then  $\vfi^{th_n}(1)$
converges to 1, for all $t\in \mathbb{R}$. Therefore, for each
$\varepsilon>0$, $\Phi_\vfi(\frac{h_n}{\varepsilon})<1$ for large enough $n$,
that is, $\|h_n\|_\vfi\to 0$. The converse follows from Theorem
\ref{thm:perturb}.
\qed

In particular, if $h_n\in \iM_s$ is a sequence converging strongly to $h$, then
$h_n$ converges to $h$ in $\|\cdot\|_\vfi$, see \cite{ohypetz}.

\section{The dual spaces.}\label{sec:dual}

The dual space $\iM_{s,0}^*$ is obtained as the quotient space
$\iM_s^*/\{\vfi\}$. Each equivalence class in $\iM_{s,0}^*$ can be
identified with its unique element $v$ satisfying $v(1)=0$. By
Proposition \ref{prop:interior}, we have
$B_{\vfi,0}^*=B_{\Phi^*_{\vfi,0}} \sqsubseteq \iM_{s,0}^*$. By
Proposition \ref{prop:finula},  $B_{\vfi,0}^*$ is the same as
$B^*_\vfi/\{\vfi\}$.

\begin{lemma}\label{lemma:conjugate}
Let $\bar c_\vfi$ be the restriction of $c_\vfi$ to $B_{\vfi,0}$. Then
the conjugate functional is $\bar c^*_\vfi(v)=F_\vfi(v+\vfi)$.
\end{lemma}

{\it Proof.} Let $v\in B^*_\vfi$, $v(1)=0$. Then by (\ref{eq:cadit}),
\begin{eqnarray*}
F_\vfi(v+\vfi)&=&\sup_{h\in B_\vfi}\{v(h)+\vfi(h)-c_\vfi(h)\}=\\
        &=&\sup_{h\in B_\vfi}\{ v(h-\vfi(h))-\bar c_\vfi(h-\vfi(h))\}=\bar c^*_\vfi(v).
\end{eqnarray*}
\qed

Let $V$ be a Banach space and $V^*$ its dual. For any subset  $D\subset V$,
let $D^\circ$ be the polar of $D$
in $V^*$, that is, $D^\circ=\{v\in V^*,\ v(h)\le 1,\ \forall h\in
D\}$. We will need the following lemma.

\begin{lemma}\label{lemma:polar} Let
$F:V\to \mathbb{R}^+$ be a convex functional such that $F(0)=0$ and let $F^*$ be its
conjugate.  Let $D=\{x\in V,\ F(x)\le 1\}$ and $D^*=\{v\in V^*,\ F^*(v)\le 1\}$.  Then
$$
\frac 12 D^*\subseteq D^\circ\subseteq D^*
$$
\end{lemma}

{\it Proof.}
If  $v\in D^*$, then $v(x)\le F(x)+F^*(v)\le 2$ for all $x\in D$ and
therefore $\frac12v\in D^\circ$. Let now $v\in D^\circ$, then
$$
v(x)-1\le 0\le F(x)\quad \mbox{ for } x\in D
$$
If $F(x)>1$, then by continuity there is some $t\in (0,1)$ such that $F(tx)=1$.
Since $tx\in D$, $v(tx)\le 1$, moreover,
 by convexity, $1=F(tx)\le tF(x)$. Consequently,
$$
v(x)-1\le\frac1t-1\le F(x)
$$
It follows that $F^*(v)\le 1$ and $v\in D^*$.
\qed

Let us denote  $K_{\vfi,0}:=\{h\in B_{\vfi,0},\ \Phi_{\vfi,0}(h)\le 1\}$.
Then $K_{\vfi,0}$ is  the closed unit ball in $B_{\vfi,0}$.
Its polar $K^\circ_{\vfi,0}$ is the closed unit ball in $B^*_{\vfi,0}$.
\begin{prop}\label{prop:unitball} Let  $v$ be an element in $K^\circ_{\vfi,0}$.
Then there are states $\omega_1$, $\omega_2$, satisfying
$S(\omega_1,\vfi)+S(\omega_2,\vfi)\le 1$, such that
  $v=\omega_1-\omega_2$.
\end{prop}

{\it Proof.} Since $\bar c_\vfi$ is continuous on $B_{\vfi,0}$, the set
 $D:=\{ h\in B_{\vfi,0},\ \bar c_\vfi(h)\le1\}$ is closed. Let us endow the dual
pair $B_{\vfi,0}$ and
 $B_{\vfi,0}^*$ with the $\sigma(B_{\vfi,0},B_{\vfi,0}^*)$ and
 $\sigma(B_{\vfi,0}^*,B_{\vfi,0})$ topology, respectively. As $D$ is convex, it
is closed also in this weaker topology.
The set $D\cap -D$ is absolutely convex and closed, moreover,
\begin{equation}\label{eq:D_D}
D\cap-D\subseteq K_{\vfi,0}  \subseteq 2(D\cap-D),
\end{equation}
as can be easily checked.  Then
$$
\frac12 (D\cap-D)^\circ\subseteq K_{\vfi,0}^\circ\subseteq (D\cap-D)^\circ
$$
 By \cite{kothe}, $(D\cap -D)^\circ$ is the closed
convex cover of $D^\circ\cup -D^\circ$, which is the same as the
closed absolutely convex cover of $D^\circ$.
Moreover, since $D^\circ$ is the polar of a neighborhood of 0, it is
compact (\cite{kothe}). Therefore its absolutely  convex
cover is also compact, hence closed. It follows that
 $(D\cap-D)^\circ$ is the absolutely  convex cover of $D^\circ$.

By Lemma \ref{lemma:conjugate} and \ref{lemma:polar},
$$
\frac12(\iS_1-\vfi)\subseteq D^\circ\subseteq \iS_1-\vfi
$$
and this implies
\begin{equation}\label{eq:unitball}
\frac14{\rm abs}\ {\rm conv}\, (\iS_1-\vfi)\subseteq K^\circ_{\vfi,0}
\subseteq {\rm abs}\ {\rm conv}\, (\iS_1-\vfi)
\end{equation}

Let now $v\in{\rm abs}\ {\rm conv}\, (\iS_1-\vfi) $,
then there are elements $\vfi_1\dots,\vfi_n\in \iS_1$,
and real numbers $\lambda_1,\dots,\lambda_n$, $\sum_n|\lambda_n|=1$,
such that $v=\sum_n\lambda_n(\vfi_n-\vfi)$. Let $m\le n$ be such that
$\lambda_i>0$ for $i\le m$ and $\lambda_i<0$ for
$i>m$. Then $v=\omega_1-\omega_2$, with
$$
\omega_1=\sum_{i=1}^m\lambda_i\vfi_i+(1-\lambda)\vfi,\quad
\omega_2=\sum_{i=m+1}^n|\lambda_i|\vfi_i+\lambda\vfi
$$
where $\lambda=\sum_{i=1}^m\lambda_i$, moreover,
$S(\omega_1,\vfi)\le\sum_i^m\lambda_iS(\vfi_i,\vfi)\le \lambda$, and
similarly, $S(\omega_2,\vfi)\le 1- \lambda$.\qed

\begin{thm}\label{thm:duals}
\begin{enumerate}
\item [(i)] $B_\vfi^*=\iP_\vfi-\iP_\vfi$ and
$B_\vfi^*\cap \iM_*^+=\iP_\vfi$.
\item [(ii)]  $B_{\vfi,0}^*=\cup_n n(\iS_1-\iS_1)$.
\end{enumerate}
\end{thm}

{\it Proof.}
(i) Let $\omega\in B_\vfi^*$ and let $v=\omega-\omega(1)\vfi$. Then $v$ can
be seen as an element in $B^*_{\vfi,0}$. Let $\|v\|_{\vfi,0}^*=t$, then by
Proposition \ref{prop:unitball}, there are $\omega_1,\omega_2\in \iS_1$, such
that $\frac vt=\omega_1-\omega_2$, that is,
$\omega=t\omega_1+\omega(1)\vfi-t\omega_2$.
Since $\omega_1,\omega_2,\vfi\in \iP_\vfi$ and $\iP_\vfi$ is a convex cone,
it follows that $B_\vfi^*\subseteq
\iP_\vfi-\iP_\vfi$. On the other hand,
we have already shown that if $\omega\in \iS_\vfi$, then $\omega\in B_\vfi^*$
and hence $\iP_\vfi-\iP_\vfi\subseteq B_\vfi^*$.
Let $\omega\in B^*_\vfi\cap\iM_*^+$,
then we get $\omega+t\omega_2=t\omega_1+\omega(1)\vfi$. It follows that
$\omega+t\omega_2\in \iP_\vfi$, and Donald's identity implies that $\omega$ must
be in $\iP_\vfi$.

(ii) By Proposition \ref{prop:unitball},
$$
K_{\vfi,0}^\circ\subseteq(\iS_1-\iS_1)\subseteq 4K_{\vfi,0}^\circ.
$$
The equality now follows
from the fact that the closed unit ball is absorbing in
$B_{\vfi,0}^*$. \qed

In the rest of this section, we find an  equivalent norm on
$B_{\vfi,0}^*$.

We define a function $f:\ \mathfrak{S}_*\times\mathfrak{S}_*\to
\mathbb{R}^+$ by
$$
f(\omega_1,\omega_2)=S(\omega_1,\vfi)+S(\omega_2,\vfi).
$$
Clearly, $f$ is weakly lower semicontinuous and strictly convex. Further,
let $v\in \mathfrak{S}_*-\mathfrak{S}_*$ and let  $L_v=\{(\omega_1,\omega_2)\in
\mathfrak{S}_*\times\mathfrak{S}_*,\ \omega_1-\omega_2=v\}$. Then $L_v$ is a
weakly closed subset in $\iM_*\times\iM_*$.

\begin{lemma} Let $v\in \iS_\vfi-\iS_\vfi$. Then the function $f$
attains its minimum over $L_v$ at  a unique point $(v_+,v_-)\in L_v$.
\end{lemma}

{\it Proof.} By assumptions, $v=\omega_1-\omega_2$ for some
$\omega_1,\omega_2\in \iS_\vfi$. Let $C>0$ be such that $\omega_1,\omega_2\in
\iS_C$, then the infimum is taken over the set $L_v\cap \iS_C\times\iS_C$. Since
$L_v$ is weakly closed and $\iS_C$ is weakly compact, the intersection  is weakly compact
and $f$ attains its minimum on it.
 Uniqueness follows by strict convexity of $f$.
\qed

Let us now define the functional $\Psi_{\vfi,0}:\ \iM_{s,0}^*\to
\mathbb{R}^+$ by
$$
\Psi_{\vfi,0}(v)=\left \{ \begin{array}{cc} f(v_+,v_-)& \mbox{if }\ v\in
\iS_\vfi-\iS_\vfi\\
\infty &\mbox{otherwise} \end{array} \right.
$$

\begin{lemma}
$\Psi_{\vfi,0}$ is a Young function.
\end{lemma}

{\it Proof.}
It is easy to check that $\Psi_{\vfi,0}$ is  convex, positive,
$\Psi_{\vfi,0}(v)= \Psi_{\vfi,0}(-v)$ and that $\Psi_{\vfi,0}(v)=0$
if and only if $v=0$.  We will
show that $\Psi_{\vfi,0}$ is  lower semicontinuous.

To do this, we have to prove that for any $C>0$, the set of all $v$ satisfying
$\Psi_{\vfi,0}(v)\le C$ is  closed.
Let $v_n$ be a sequence of elements in this set, converging  to some $v$.
Let $v_n=v_{n+}-v_{n-}$ be the corresponding decompositions, then
$v_{n+},v_{n-}\in \iS_C$ for all $n$, hence there are elements $v'_+$ and $v'_-$
in $\iS_C$ and a subsequence
$v_{n_k}=v_{n_k+}-v_{n_k-}$ such that $v_{n_k+}\to v'_+$ and $v_{n_k-}\to v'_-$
weakly. It follows that $v=v'_+-v_-'$ and $\Psi_{\vfi,0}(v)\le
S(v'_+,\vfi)+S(v'_-,\vfi)\le \liminf S(v_{n_k+},\vfi)+S(v_{n_k-},\vfi)\le C$.

Suppose that $v\ne 0$, then $\Psi_{\vfi,0}(v)>0$.
If $t>1$, then
by convexity, $t\Psi_{\vfi,0}(v)\le \Psi_{\vfi,0}(tv)$, hence
$\lim_{t\to\infty} \Psi_{\vfi,0}(tv)=\infty$.
\qed

Let us find the corresponding Banach space. Note that
$$
C_{\Psi_{\vfi,0}}=\{ \omega_1-\omega_2\ :\ \omega_1,\omega_2\in \mathfrak S_*,
S(\omega_1,\vfi)+S(\omega_2,\vfi)\le1\}.
$$
By Proposition \ref{prop:unitball},
this implies that
$K_{\vfi,0}^\circ\subseteq  C_{\Psi_{\vfi,0}}\subseteq \iS_1-\iS_1$
and by Theorem \ref{thm:duals} (ii),
$B_{\vfi,0}^*\subseteq L_{\Psi_{\vfi,0}}\subseteq B_{\vfi,0}^*$.

\begin{prop}   $\|\cdot\|_{\Psi_{\vfi,0}}$ defines an  equivalent norm in
$B^*_{\vfi,0}$.
\end{prop}

{\it Proof.} Let $\Psi_{\vfi,0}^*:\iM_s\to\mathbb{R}$ be the conjugate
functional, then
\begin{eqnarray*}
\Psi_{\vfi,0}^*(h)&=&\sup_{v\in \iM_{s,0}^*}v(h)-\Psi_{\vfi,0}(v)=\\
&=&\sup_{v\in \iS_\vfi-\iS_\vfi}\sup_{(\omega_1,\omega_2)\in L_v}
\omega_1(h)-\omega_2(h)-f(\omega_1,\omega_2)=\\ &=&\sup_{\omega_1,\omega_2\in
\iS_\vfi}\omega_1(h)-S(\omega_1,\vfi)+\omega_2(-h)-S(\omega_2,\vfi)=
2\Phi_{\vfi,0}(h)
\end{eqnarray*}
It follows that $\Psi_{\vfi,0}(v)=\Psi_{\vfi,0}^{**}(v)=2\Phi_{\vfi,0}^*(\frac
12 v)$.
Since the norms $\|\cdot\|_{\vfi,0}^*$ and $\|\cdot\|_{\Phi^*_{\vfi,0}}$ are
equivalent, this finishes the proof.
\qed

\section{The chain rule.}

\begin{prop}\label{prop:chain}  Let $h \in B_\vfi$, $k\in \iM_s$. Then
$[\vfi^{h+k}]=[[\vfi^h]^k]$, $c_\vfi(h+k)=c_{[\vfi^h]}(k)+c_\vfi(h)$ and for all
normal states $\omega$ the equality
\begin{equation}\label{eq:chain}
\omega(k)-S(\omega,[\vfi^h])=c_\vfi(h+k)-c_\vfi(h)-S(\omega,[\vfi^{h+k}])
\end{equation}
holds.
\end{prop}

{\it Proof.}
Let $h_n\in \iM_s$ be such that $h_n\to h$ in $B_\vfi$.
By  the chain rule (\ref{eq:chainM}), we have
$[\vfi^{h_n+k}]=[[\vfi^{h_n}]^k]$ and
$c_\vfi(h_n+k)=c_{[\vfi^{h_n}]}(k)+c_\vfi(h_n)$.
By Theorem \ref{thm:perturb}, $c_\vfi(h_n)\to c_\vfi(h)$,
$c_\vfi(h_n+k)\to c_\vfi(h+k)$ and
$[\vfi^{h_n}]\to[\vfi^h]$, $[\vfi^{h_n+k}]\to[\vfi^{h+k}]$ strongly.
Now we can proceed exactly as in the proof of Theorem 12.10 in \cite{ohypetz}
to obtain
(\ref{eq:chain}). By putting $\omega=[\vfi^{h+k}]$ in this equality,
we get
$$
[\vfi^{h+k}](k)+S([\vfi^{h+k}],[\vfi^h])=c_\vfi(h+k)-c_\vfi(h)\ge
\omega(k)-S(\omega,[\vfi^h]),
$$
for all $\omega$,
which implies the statement of the proposition.
\qed

\begin{thm}\label{thm:eqiuv}
Let  $h\in B_\vfi$. Then $B_{[\vfi^h]}=B_\vfi$ and $\iS_{[\vfi^h]}=\iS_\vfi$.
\end{thm}

{\it Proof.}
Let $k\in \iM_s$ and let $\varepsilon >0$. By Proposition
\ref{prop:chain},
$$c_{[\vfi^h]}(k)=c_\vfi(h+k)-c_\vfi(h).$$
Since $c_\vfi$ is  continuous on $B_\vfi$, there is a   $\delta>0$
such that
$$
|c_\vfi(h+k)-c_\vfi(h)|<\log 2
$$
if $\|k\|_\vfi<\delta$. It follows that
$\|k\|_{[\vfi^h]}<\varepsilon$ whenever $\| k\|_\vfi<\delta\varepsilon$
and this implies $B_\vfi\sqsubseteq B_{[\vfi^h]}$.
In particular,
$h\in B_{[\vfi^h]}$.

Let $h_n$ be a sequence converging to $h$ in $B_\vfi$, then by (\ref{eq:sup2})
$$
\omega(h_n)-S(\omega,\vfi)=c_\vfi(h_n)-S(\omega,[\vfi^{h_n}])
$$
By Theorem \ref{thm:perturb}, and lower semicontinuity,
$$
\omega(h)-S(\omega,\vfi)\le c_\vfi(h)-S(\omega,[\vfi^h])
$$
This implies $\iS_\vfi\subseteq \iS_{[\vfi^h]}$.

Further, $h_n$ converges to  $h$ in $B_{[\vfi^h]}$ and
by Theorem \ref{thm:perturb} and Proposition \ref{prop:chain},
$$
[[\vfi^h]^{-h}]=\lim_n [[\vfi^h]^{-h_n}]=
\lim_n[\vfi^{h-h_n}]=\vfi.
$$
By the first part of the proof, $B_{[\vfi^h]}=B_\vfi$ and $\iS_\vfi=
\iS_{[\vfi^h]}$.
\qed

\begin{thm}\label{thm:chain} Let $h,k\in B_\vfi$. Then the chain rule
$c_\vfi(h+k)=c_{[\vfi^h]}(k)+c_\vfi(h)$, $[[\vfi^h]^k]=[\vfi^{h+k}]$ holds.
\end{thm}

{\it Proof.} Let $k_n\in \iM_s$ be a sequence converging to
$k$ in $B_\vfi=B_{[\vfi^h]}$. Then
$$
[[\vfi^h]^{k}]=\lim_n [[\vfi^h]^{k_n}]=
\lim_n[\vfi^{h+k_n}]=[\vfi^{h+k}].
$$
and by Proposition \ref{prop:chain},
$$
c_\vfi(h+k)=\lim_n c_{[\vfi^h]}(k_n)+c_\vfi(h)=c_{[\vfi^h]}(k)+
c_\vfi(h)
$$

\qed

\begin{coro}\label{coro:equality}
Let $h\in B_\vfi$ and let $\omega$ be a normal state. Then
the equality
$$
\omega(h)-S(\omega,\vfi)=c_\vfi(h)-S(\omega,[\vfi^h])
$$
holds.
\end{coro}

{\it Proof.} By (\ref{eq:sup2}) and lower semicontinuity, we have
$$
\omega(h)-S(\omega,\vfi)\le c_\vfi(h)-S(\omega,[\vfi^h])
$$
Since, by the chain rule, $\vfi=[[\vfi^h]^{-h}]$ and
$c_{[\vfi^h]}(-h)=-c_\vfi(h)$, we also have
$$
\omega(-h)-S(\omega,[\vfi^h])\le
c_{[\vfi^h]}(-h)-S(\omega,\vfi)=-c_\vfi(h)-S(\omega,\vfi)
$$
which implies the opposite inequality.
\qed

\begin{coro}\label{coro:unique}
Let $[\vfi^h]=[\vfi^k]$ for some $h,k\in B_\vfi$. Then  $h-k=\vfi(h-k)$.

\end{coro}

{\it Proof.} Let us suppose that $h\in B_\vfi$ is such that
$[\vfi^h]=\vfi$. Then $[\vfi^{nh}]=\vfi$ for all $n\in {\mathbb{N}}$.
It follows that  $c_\vfi(nh)=n\vfi(h)=nc_\vfi(h)$ for all $n$ and
for  $0\le t\le 1$, we have by (\ref{eq:gt}) and convexity of $c_\vfi$ that
$$
tc_\vfi(h)=\vfi(th)\le c_\vfi(th)\le tc_\vfi(h)
$$
It follows that $c_\vfi(th)=tc_\vfi(h)=t\vfi(h)$ for all $t\ge 0$. Since
also $[\vfi^{-h}]=[[\vfi^h]^{-h}]=\vfi$, we have
$c_\vfi(-th)=tc_\vfi(-h)=-t\vfi(h)$ for $t\ge 0$.

It is easy to see that $c_\vfi(k-\lambda)=c_\vfi(k)-\lambda$ for all
$k\in B_\vfi$ and $\lambda\in\mathbb{R}$. Let $\lambda=\vfi(h)$, then it
follows that
$$
c_\vfi(t(h-\lambda))=0=c_\vfi(t(-h+\lambda))
$$
for all $t\ge 0$. This implies $\|h-\lambda\|_\vfi=0$ and hence $h=\lambda$.

 Let now $[\vfi^h]=[\vfi^k]$, then $[[\vfi^k]^{-h}]=[\vfi^{k-h}]=
\vfi$ and $h-k=\lambda=\vfi(h-k)$.
\qed

Note that the function $\bar c_\vfi: B_{\vfi,0}\to\mathbb R$ corresponds to the cumulant generating
functional in the commutative case. Let us list some of its properties.

\begin{thm}\label{thm:cumulant}
The function $\bar c_\vfi$ has the following properties.
\begin{enumerate}
\item[(i)] $\bar c_\vfi$ is positive, strictly convex and continuous, $\bar c_\vfi(0)=0$.
\item[(ii)] $\bar c_\vfi$ is Gateaux differentiable, with  $\bar c'_\vfi(h)=[\vfi^h]-\vfi$
\item[(iii)] The map
$$B_{\vfi,0}\ni h\mapsto [\vfi^h]-\vfi\in B^*_{\vfi,0}$$
is one-to-one
and norm to
$\sigma(B^*_{\vfi,0},B_{\vfi,0})$-continuous.
\end{enumerate}
\end{thm}

{\it Proof.} (i) By Corollary \ref{coro:equality},
$\bar c_\vfi(h)=S(\vfi,[\vfi^h])\ge 0$ and $\bar c_\vfi(0)=0$ by definition. Let now
$h,k\in B_{\vfi,0}$ and $0<\lambda< 1$ be such that
$$\bar c_\vfi(\lambda h+(1-\lambda)k)=\lambda \bar c_\vfi(h)+(1-\lambda)\bar c_\vfi(k).$$
Then
\begin{eqnarray*}
&\sup_{\iS_\vfi}& \lambda(\omega(h)-S(\omega,\vfi))+(1-\lambda)(\omega(k)-S(\omega,\vfi))
= \\
&=& \lambda \sup_{\iS_\vfi} (\omega(h)-S(\omega,\vfi))
+(1-\lambda)\sup_{\iS_\vfi}(\omega(k)-S(\omega,\vfi))
\end{eqnarray*}
This implies that the maximum in both expressions on the right hand side
is attained at the same point.
Therefore  $[\vfi^h]=[\vfi^k]$, hence $h-k=\vfi(h-k)=0$.

(ii) By Theorem \ref{thm:perturb}, $[\vfi^h]-\vfi$ is the unique element in
$B^*_{\vfi,0}$, such that
$$([\vfi^h]-\vfi)(h)=\bar c_\vfi(h)+\bar c^*_\vfi([\vfi^h]-\vfi).
$$
By \cite{ektem}, this implies that $\bar c_\vfi$ is Gateaux differentiable in $h$
with derivative $\bar c'_\vfi(h)=[\vfi^h]-\vfi$.

(iii) Let $h_n\to h$ in $B_\vfi$, then
 $[\vfi^{h_n}]$
 converges strongly to $[\vfi^h]$ and $S([\vfi^{h_n}],\vfi)\to
 S([\vfi^h],\vfi)$.
 It follows that $[\vfi^{h_n}](k)\to [\vfi^h](k)$ for each $k\in \iM_s$ and
 moreover, the set $\{[\vfi^{h_n}], n\in \mathbb N\}$ is equicontinuous in $B_\vfi^*$. This implies that
 $[\vfi^{h_n}](k)\to [\vfi^h](k)$ for all $k\in B_\vfi$. The map is one-to-one by Corollary \ref{coro:unique}.

\qed

 \section{A manifold structure on faithful states.}\label{sec:mani}

Recall that a $C^p$-atlas on a set $X$ is a family of pairs
$\{(U_i,e_i)\}$, such that
\begin{enumerate}
\item[(i)] $U_i\subset X$ for all $i$ and $\cup U_i=X$.
\item[(ii)] For all $i$,  $e_i$ is a bijection of $U_i$ onto an open subset
$e_i(U_i)$ in some Banach space $B_i$, and for  $i,j$, $e_i(U_i\cap U_j)$ is
open in $B_i$.
\item[(iii)] The map $e_je_i^{-1}:\ e_i(U_i\cap U_j)\to e_j(U_i\cap U_j)$
is a $C^p$- isomorphism for all $i,j$.

\end{enumerate}

Let $\mathcal F_*$ be the set of faithful normal states on $\iM$.
For $\vfi\in \mathcal F_*$, let $V_\vfi$ be the open unit ball in $B_{\vfi,0}$ and
let $s_\vfi: V_\vfi\to \mathcal F_*$ be the map $h\mapsto [\vfi^h]$.
By Corollary
\ref{coro:unique}, $s_\vfi$ is a bijection onto the set
$s_\vfi(V_\vfi)=:U_\vfi\subset \mathcal S_\vfi$. Let  $e_\vfi$ be the
restriction of $s^{-1}_\vfi$ to $U_\vfi$. Then we have

 \begin{thm} $\{(U_\vfi, e_\vfi), \vfi\in \mathcal F_*\}$ is a
 $C^{\infty}$-atlas on $\mathcal F_*$.
\end{thm}

{\it Proof.} The property (i) and the first part of (ii)
of the definition of the $C^p$ atlas are obviously
satisfied. Let $\vfi_1,\vfi_2\in \mathcal F_*$ be such that
$U_{\vfi_1}\cap U_{\vfi_2}\ne \emptyset$. We prove that
$e_{\vfi_1}(U_{\vfi_1}\cap U_{\vfi_2})$ is open in $B_{\vfi_1,0}$.

Let
$h_1\in e_{\vfi_1}(U_{\vfi_1}\cap
U_{\vfi_2})$. Then
there is some   $h_2\in B_{\vfi_2,0}$, such that
$[\vfi_1^{h_1}]=[\vfi_2^{h_2}]$.
By Theorem \ref{thm:eqiuv}, $B_{\vfi_1}=B_{[\vfi_1^{h_1}]}=B_{[\vfi_2^{h_2}]}=
B_{\vfi_2}$ and by the chain rule, $\vfi_1=[\vfi_2^k]$, where
$k=h_2-h_1+\vfi_2(h_1)\in B_{\vfi_2,0}$. Clearly, the map
$B_{\vfi_1,0}\to B_{\vfi_2,0}$, given by $h\mapsto
h-\vfi_2(h)$ is continuous.

Let $\varepsilon>0$ be such that $h_2+h_2'\in V_{\vfi_2}$ whenever
$\|h'_2\|_{\vfi_2}<\varepsilon$ and let us choose  $\delta>0$ such that
$h_1+h_1'\in V_{\vfi_1}$ and $\|h'_1-\vfi_2(h_1')\|_{\vfi_2}< \varepsilon$ for
$\|h_1'\|_{\vfi_1}<\delta$. For such $h_1'$, we have
$$
s_{\vfi_1}(h_1+h_1')=[\vfi_1^{h_1+h_1'}]=[\vfi_2^{k+h_1+h_1'-\vfi_2(h_1')}]=
[\vfi_2^{h_2+h_1'-\vfi_2(h_1')}]\in U_{\vfi_1}\cap U_{\vfi_2}
$$
This proves that $s_{\vfi_1}^{-1}(U_{\vfi_1}\cap U_{\vfi_2})$ is open in
$B_{\vfi_1,0}$. It is also clear that the map
\begin{eqnarray*}
s^{-1}_{\vfi_2}s_{\vfi_1}: s^{-1}_{\vfi_1}(U_{\vfi_1}\cap U_{\vfi_2})&\to&
s^{-1}_{\vfi_2}(U_{\vfi_1}\cap U_{\vfi_2})\\
h &\mapsto& k+h-\vfi_2(h)
\end{eqnarray*}
is $C^\infty$, which proves (iii).
\qed

It is not difficult to see that for $\vfi\in \mathcal F_*$,
the set $\mathcal F_\vfi:=\{[\vfi^h],\ h\in B_{\vfi,0}\}$
is a connected component of the manifold.
Let us now define a family of mappings
$$
U^{(e)}_{\vfi_1,\vfi_2}: B_{\vfi_1,0}\ni h\mapsto h-\vfi_2(h)\in B_{\vfi_2,0},\quad \vfi_1,\vfi_2\in
\mathcal F_\vfi
$$
It is clear that this defines a parallel transport on the tangent bundle
of $\mathcal F_\phi$ and the associated
 globally flat affine connection is  the exponential connection, \cite{gipi}.

Let us recall that the dual connection is defined on the cotangent bundle $T^*\mathcal F_\vfi$ by means of
the parallel transport $\{ (U^{(e)}_{\vfi_2,\vfi_1})^*,\ \vfi_1,\vfi_2\in \mathcal F_\vfi\}$, where
$$
 \langle (U^{(e)}_{\vfi_2,\vfi_1})^*v,h\rangle = \langle v,
U^{(e)}_{\vfi_1,\vfi_2}h\rangle, \quad v\in B_{\vfi_2,0}^*, h\in B_{\vfi_1,0},
$$
and the duality is given by $\langle v, h\rangle = v(h)$. Since $v(h-\vfi_1(h))=v(h)$ for all $\vfi_1$,
the dual parallel transport is
$$
U^{(m)}_{\vfi_1,\vfi_2}: B_{\vfi_1,0}^*\ni v\mapsto v\in B_{\vfi_2,0}^*,\quad \vfi_1,\vfi_2\in
\mathcal F_\vfi
$$
which corresponds to the mixture connection.

\end{document}